\documentclass{llncs}
\usepackage[utf8]{inputenc}
\usepackage[T1]{fontenc}
\usepackage{graphicx}
\usepackage{amsmath}
\usepackage{cite}
\usepackage{multirow}
\usepackage{xcolor}
\usepackage{hyperref}
\usepackage{tabularx}
\usepackage{makecell}
\usepackage{footnote}
\makesavenoteenv{tabular}
\makesavenoteenv{table}

\newcommand{\ie}{\emph{i.e.},~}
\newcommand{\eg}{\emph{e.g.},~}
\newcommand{\ea}{\emph{et al.}~}
\newcommand{\vs}{\emph{vs.}~}

\author{Huy-Dung Nguyen \and Michaël Clément \and Boris Mansencal \and Pierrick Coupé}
\institute{Univ. Bordeaux, CNRS, Bordeaux INP, LaBRI, UMR 5800, 33400 Talence, France\\
\email{huy-dung.nguyen@u-bordeaux.fr}
}

\title{Interpretable differential diagnosis for Alzheimer’s disease and Frontotemporal dementia}

\begin{document}

\maketitle

\begin{abstract}

Alzheimer’s disease and Frontotemporal dementia are two major types of dementia. Their accurate diagnosis and differentiation is crucial for determining specific intervention and treatment. However, differential diagnosis of these two types of dementia remains difficult at the early stage of disease due to similar patterns of clinical symptoms. Therefore, the automatic classification of multiple types of dementia has an important clinical value. So far, this challenge has not been actively explored. Recent development of deep learning in the field of medical image has demonstrated high performance for various classification tasks. In this paper, we propose to take advantage of two types of biomarkers: structure grading and structure atrophy. To this end, we propose first to train a large ensemble of 3D U-Nets to locally discriminate healthy versus dementia anatomical patterns. The result of these models is an interpretable 3D grading map capable of indicating abnormal brain regions. This map can also be exploited in various classification tasks using graph convolutional neural network. Finally, we propose to combine deep grading and atrophy-based classifications to improve dementia type discrimination. The proposed framework showed competitive performance compared to state-of-the-art methods for different tasks of disease detection and differential diagnosis.

\end{abstract}

\section{Introduction}

Alzheimer’s disease (AD) and Frontotemporal dementia (FTD) are the first and third leading causes of early-onset dementia \cite{bang_frontotemporal_2015}. The detection of these diseases is critical for the development of novel therapies. Besides, people with FTD are often misdiagnosed with AD, although these diseases have different clinical diagnostic criteria \cite{rascovsky_sensitivity_2011, mckhann_diagnosis_2011},  due to similar clinical symptoms such as a behavior or language disorder \cite{alladi_focal_2007} and brain atrophy \cite{neary_frontotemporal_2005}. This is especially true for behavioral variant of FTD (bvFTD) which is the most common variant of FTD \cite{rascovsky_sensitivity_2011}. Indeed, many studies have shown that cognitive tests fail to accurately identify FTD from AD population \cite{hutchinson_neuropsychological_2007, yew_lost_2013}. This raises the need for an early and accurate differential diagnosis to determine specific intervention and slow down the disease progression. Consequently, a multi-class differential diagnosis method – able to differentiate AD, FTD and cognitively normal (CN) subjects – would be a highly valuable tool in clinical practice. Indeed, such tool can assist clinicians in making more informed decision in a general context.

The atrophy patterns of AD and FTD can be identified with the help of structural magnetic resonance imaging (sMRI) \cite{du_different_2006, moller_alzheimer_2016}. Moreover, the affected regions may be different between diseases \cite{davatzikos_individual_2008}. Hence, it should be beneficial to use sMRI for disease detection and differential diagnosis. In the past, some methods have been proposed for this problem using volumetric and shape measurements obtained from sMRI \cite{du_different_2006, rabinovici_distinct_2008}. However, the large majority of existing methods considered only binary classification problems (\eg AD \vs CN, FTD \vs CN, FTD \vs AD). For the multi-class differential diagnosis, only a few works have been proposed \cite{bron_multiparametric_2017, ma_differential_2020, hu_deep_2021}. Moreover, the majority of existing approaches in this domain used traditional machine learning methods based on handcrafted features that may not fully exploit the image information. Therefore, Deep learning methods have been recently studied. However, the results of these approaches are usually not easily interpretable. This limits our understanding about the disease patterns.

In this paper, we propose a new method to perform specific-disease diagnosis (\ie AD \vs CN and FTD \vs CN) and differential diagnosis (\ie AD \vs FTD and AD \vs FTD \vs CN). Our purpose is to expand the knowledge about dementia sub-types and to offer an accurate diagnosis tool in a real clinical scenario. To this end, our contributions are twofold. First, we generate a 3D grading map reflecting the abnormality level of brain structures. This interpretable biomarker may assist clinicians in localizing the abnormal regions of brain, allowing a deeper understanding about multiple disease signatures. To do so, we extend the recently proposed Deep Grading (DG) framework \cite{nguyen_deep_2021} by training it with multiple types of dementia (\ie AD and FTD). Then, we classify these DG features using a graph convolutional neural network (GCN) \cite{kipf_semi-supervised_2017} to better capture disease signatures. Second, we propose to ensemble the GCN decision with a support vector machine (SVM) using brain structure volumes to improve differential diagnosis accuracy. By combining structure grading and structure atrophy, the proposed framework offers state-of-the-art performance in both disease detection and differential diagnosis.

\section{Materials and method}

\subsection{Datasets}

In this study, we used 2036 MRIs extracted from multiple open access databases: the Alzheimer's Disease Neuroimaging Initiative (ADNI) \cite{jack_alzheimers_2008}, the Open Access Series of Imaging Studies (OASIS) \cite{lamontagne_oasis_3_2019}, the Australian Imaging, Biomarkers and Lifestyle (AIBL) \cite{ellis_australian_2009}, the Minimal Interval Resonance Imaging in Alzheimer's Disease (MIRIAD) \cite{malone_miriadpublic_2013} and the Frontotemporal lobar Degeneration Neuroimaging Initiative (NIFD) \footnote{Available at \url{https://ida.loni.usc.edu/}}. All the baseline T1-weighted MRIs available in these databases were used. The NIFD dataset contains FTD patients and CN subjects while other datasets contain AD patients and CN subjects. In NIFD dataset, we only used the behavioral variant sub-type (bvFTD) which is the most prevalent form of FTD. We use all data available in ADNI1 (\ie 191 AD and 191 CN) and 90 subjects from NIFD (\ie 45 FTD, 45 CN) for training and validation. We use stratified splitting strategy to obtain 80\% training and 20\% validation data.
To eliminate possible biases during training, the same number of patients and healthy people with no significant difference in age distribution were chosen.
The other subjects (\ie 1199 CN, 371 AD and 29 FTD) were used only at the final evaluation. Table~\ref{table:datasets} describes the number of participants of each dataset used in this study.

\begin{table}[t]
    \centering
    \caption{Number of participants used in our study.}
    \label{table:datasets}
    \begin{tabular}{l@{\hskip 3em}c@{\hskip 2em}c@{\hskip 2em}c}
        \hline
        Dataset      & CN         & AD        & FTD \\
        \hline
        ADNI1        &  191       & 191       & - \\
        ADNI2        &  149       & 181       & - \\
        AIBL         &  233       & 47        & - \\
        OASIS3       &  658       & 97        & - \\
        MIRIAD       &  23        & 46        & - \\
        NIFD         &  136       & -         & 74 \\
        \hline
    \end{tabular}
\end{table}

\subsection{Preprocessing}
\label{subsection:preprocessing}
The preprocessing of the T1w MRI consisted of 5 steps: (1) denoising \cite{manjon_adaptive_2010}, (2) inhomogeneity correction \cite{tustison_n4itk_2010}, (3) affine registration into MNI152 space ($181\times217\times181$ voxels at $1mm\times1mm\times1mm$) \cite{avants_reproducible_2011}, (4) intensity standardization \cite{manjon_robust_2008} and (5) intracranial cavity (ICC) extraction \cite{manjon_nonlocal_2014}. Then, we used AssemblyNet~\footnote{\url{https://github.com/volBrain/AssemblyNet}}  \cite{coupe_assemblynet_2020} with its default hyper-parameters to segment 133 brain structures (see Figure \ref{figure:pipeline}). The brain structure segmentation was used to measure the structure volumes (\ie normalized volume in \% of ICC) and aggregate information to compute the structure grading (see Section \ref{section:method} and Figure \ref{figure:pipeline}).

\subsection{Method}
\label{section:method}

\begin{figure}[ht]
\centering
\includegraphics[width=\textwidth]{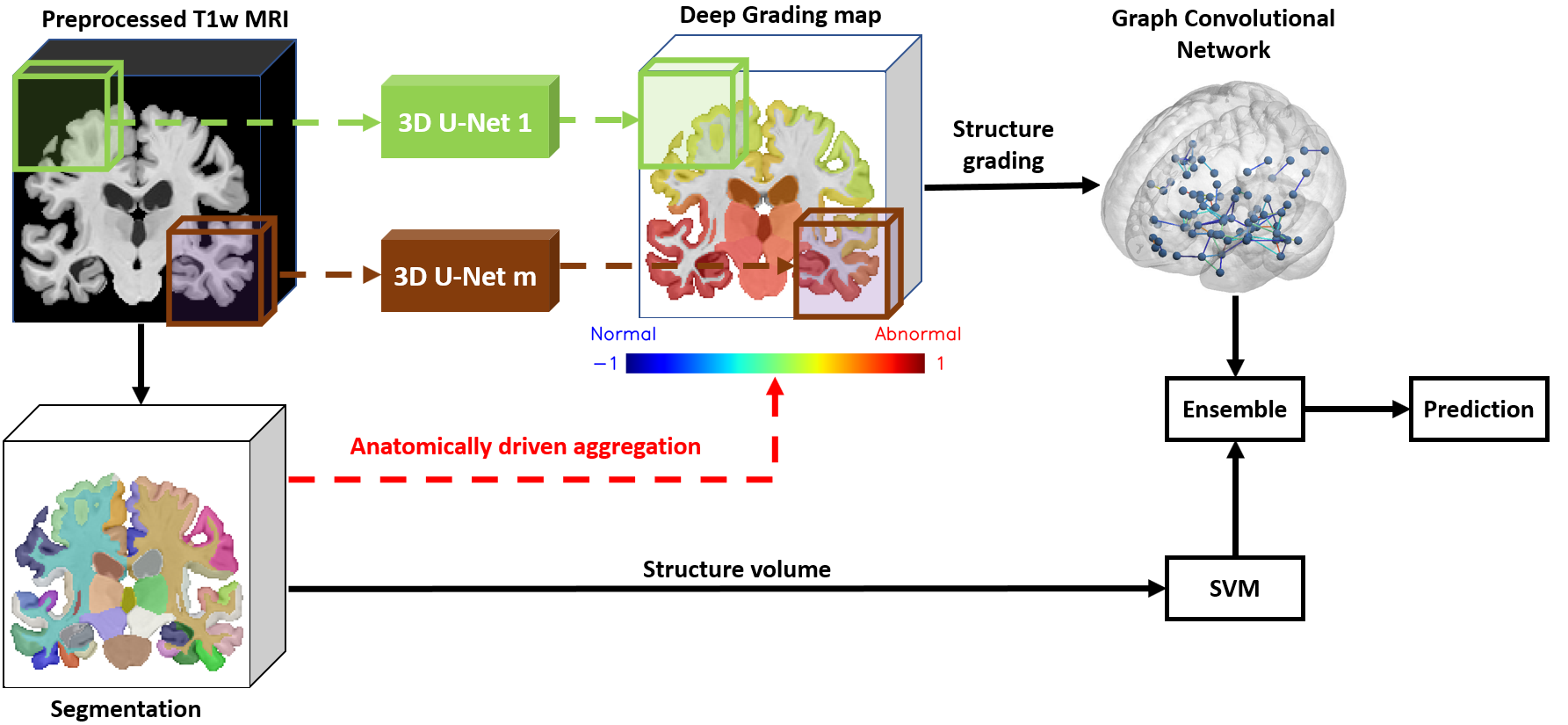}
\caption{An overview of the proposed method. The T1w image, its segmentation and the deep grading map are taken from an AD patient.}
\label{figure:pipeline}
\end{figure}

Recently, a Deep Grading (DG) framework has been proposed for AD detection as an efficient and interpretable tool \cite{nguyen_deep_2021}. Here, we proposed to extend this approach to differential diagnosis and to combine it with atrophy-based features through an ensemble strategy (see Figure~\ref{figure:pipeline}).\\
\textbf{Grading-based classification:}
First, a preprocessed MRI was downsampled with a factor of 2 to reduce the computational cost. This was used to extract $k \times k \times k$ (\ie $k=5$) overlapping patches of the same size (\ie $32 \times 48 \times 32$ voxels) uniformly distributed across the whole brain.
We used $m=k\times k\times k$ (\ie $m=125$) 3D U-Nets to grade these patches. Concretely, each of the $m$ patch locations was analyzed by one specific U-Net. This U-Net was trained to predict a 3D grading map whose each voxel reflects the degree of similarity to normal or abnormal group. The obtained grading values were assembled to reconstitute a global grading map which was interpolated to the original input size. The training procedure of deep grading part was similar to  \cite{nguyen_deep_2021}. However, we considered the brain changes of both AD and FTD as abnormal patterns instead of specific patterns of AD. Thus, for the ground-truth, we assign the value 1 (resp. $-1$) to all voxels inside a patch extracted from an AD/FTD patient (resp. CN subject). All voxels outside of ICC are set to 0.
After that, we computed $s = 133$ average grading scores (one per brain structure) using a structural segmentation (obtained with AssemblyNet \cite{coupe_assemblynet_2020}). By doing this, each subject is encoded into an s-dimensional vector. Then, we defined a fully-connected graph of $s$ nodes. Each node embeds structure grading score and subject's age. A GCN classifier with 3 layers of 32 channels was used to classify this graph. While training the GCN model, due to the imbalance nature of training set, we applied oversampling technique to balance classes in order to make the model more robust.\\
\textbf{Atrophy-based classification:}
Parallel to the deep grading model, we used the volume of the 133 brain structures as an additional feature vector to represent atrophy information. To exploit these features, we trained a linear SVM model for the same classification problem. The data used in training the SVM model was the same as training the deep grading model. We used a grid search of three kernels (linear, polynomial, and gaussian) and 500 values of the hyper-parameter C in $[10^{-5}, 10^5]$ on the validation set for tuning hyper-parameters. During training, we used balanced weights (available in scikit-learn library \cite{pedregosa11a}) to compensate for class imbalance.

Finally, we fused the probability vector of GCN and SVM by estimating their best linear combination on training data for the final decision. The training data used for the deep grading model and all the classifiers (\ie GCN, SVM or ensemble) came from the same subjects.

\section{Experimental results}
\subsection{Ablation study for binary classification tasks}

Table~\ref{table:mixing_for_binary_classification} shows the results of our ablation study dedicated to binary classification tasks. First, we propose to perform dementia diagnosis (\ie AD and FTD \vs CN), AD diagnosis (\ie AD \vs CN), FTD diagnosis (\ie FTD \vs CN) and 2-class differential diagnosis (\ie AD \vs FTD).

\noindent
\textbf{Multiple diseases training for specific disease classifications.} In this part, we assessed the influence of mixing multiple diseases in training for specific disease classifications. To do this, we trained deep grading model + classifier respectively with ADNI1 ($N = 382$), NIFD ($N = 90$) and ADNI1 $+$ NIFD ($N = 472$) (see Table \ref{table:mixing_for_binary_classification}). For AD diagnosis, we observed that mixing AD and FTD (using ADNI1 $+$ NIFD) during training can improve the model performance compared to training only on AD, CN subjects (see exp. 1 \vs 3; exp. 7 \vs 9). For FTD diagnosis, this training strategy yielded better results for all classifiers (see exp. 2 \vs 3; exp. 5 \vs 6; exp. 8 \vs 9). Overall, training on multiple diseases can improve, in most of cases, the performance compared to directly train the classifier for a specific task.

\noindent
\textbf{Combining grading and atrophy for better diagnosis.} We observed that GCN based on DG features obtained the best results for almost all the binary diagnosis tasks compared to SVM with balanced accuracies (BACC) similar or higher than 88\%.  For 2-class differential diagnosis AD \vs FTD, SVM based on atrophy features outperformed GCN and obtained 89.5\% in BACC. Consequently, we propose to combine both strategies in order to take advantage of the high capability of deep grading framework to detect pathologies and the good performance of atrophy-based classification to differentiate AD from FTD. As expected, the combined model (see exp. 9) yielded the best performance for all the considered classification tasks (see Table~\ref{table:mixing_for_binary_classification}).

\begin{table}[ht]
    \centering
    \caption{Ablation study of our method for binary classification tasks. We use the balanced accuracy (BACC) to assess the performance. The reported performances are the average of 10 repetitions and presented in \%. {\color{red} Red}: best result, {\color{blue} Blue}: second best result.}
    \label{table:mixing_for_binary_classification}
    \begin{tabular}{c@{\hskip 0.6em}l@{\hskip 0.6em}c@{\hskip 0.6em}c@{\hskip 0.6em}c@{\hskip 0.6em}c@{\hskip 0.6em}c}
        \hline
        No. & Training set   & Classifier  & \makecell{Dementia \\ diagnosis\\Dem. \vs CN\\$N=1554$} & \makecell{AD\\diagnosis\\AD \vs CN\\$N=1525$}         & \makecell{FTD\\diagnosis\\FTD \vs CN\\$N=1183$} & \makecell{Differential \\ diagnosis\\AD \vs FTD\\$N=400$}\\
        \hline
        1 & ADNI1     & GCN & -& 89.6       & - & -\\
        2 & NIFD      & GCN & -& -       & 88.0  & -\\
        3 & ADNI1$+$NIFD  & GCN &  {\color{blue} \textbf{90.3}}     &  {\color{blue} \textbf{90.1}}       & {\color{blue} \textbf{92.3}} & 74.6\\
        
        4 & ADNI1     & SVM & -& 86.3       & - & -\\
        5 & NIFD      & SVM & -& -       & 88.4  & -\\
        6 & ADNI1$+$NIFD  & SVM & 86.7 & 86.2       & 91.7 & {\color{blue} \textbf{89.5}}\\
        
        7 & ADNI1     & Ensemble & -& 89.9       & - & -\\
        8 & NIFD      & Ensemble & -& -       & 88.4  & -\\
        9 & ADNI1$+$NIFD    & Ensemble & {\color{red} \textbf{90.5}} & {\color{red} \textbf{90.3}}       & {\color{red} \textbf{93.3}} & {\color{red} \textbf{89.7}}\\
        
        \hline
    \end{tabular}
\end{table}

\subsection{Performance for multi-disease classification}

Table~\ref{table:multi_class_classification} presents the results obtained for the 3-class differential diagnosis (\ie AD \vs CN \vs FTD). We studied the performance of GCN using deep grading, SVM using brain structure volumes and the proposed combination. As previously, the proposed combination of deep grading and atrophy-based features obtained the best results. Indeed, this model yielded better performance for all of three metrics: accuracy (ACC), balanced accuracy (BACC) and area under curve (AUC). GCN obtained the best results only for FTD sensitivity. These ensembling results were then chosen to compare with current state-of-the-art methods.

\begin{table}[ht]
    \centering
    \caption{Performance of different models for the multiple disease classification. We denote "Sen." for sensitivity. The reported performances are the average of 10 repetitions and presented in \%. The best performances are in {\color{red} red}.}
    \label{table:multi_class_classification}
    \begin{tabular}{l@{\hskip 3em}c@{\hskip 1em}c@{\hskip 1em}c@{\hskip 1em}c@{\hskip 1em}c@{\hskip 1em}c}
        \hline
        Classifier    & ACC         & BACC & AUC & CN Sen. & AD Sen. & FTD Sen.\\
        \hline
        GCN           & 85.5 & 80.7 & 91.3 & 91.2 & 66.3 & {\color{red} \textbf{84.5}}\\
        SVM       & 89.3 & 83.4 & 94.8 & 92.9 & 77.9 & 76.9\\
        Ensemble  & {\color{red} \textbf{90.4}} & {\color{red} \textbf{85.4}} & {\color{red} \textbf{95.3}} & {\color{red} \textbf{93.4}}& {\color{red} \textbf{81.0}}& 81.7 \\
        \hline
    \end{tabular}
\end{table}

\subsection{Comparison with state-of-the-art methods}

Most of existing methods are based on classification using machine learning of volume-based or surface-based features. Kim \ea used a linear discriminant analysis to classify cortical thickness (Cth) for dementia diagnosis and 2-class differential diagnosis (\ie AD \vs FTD) \cite{kim_machine_2019}. Möller \ea used an SVM to classify tissue density map for binary classification tasks \cite{moller_alzheimer_2016}. In \cite{Yu_mri-based_2021}, Yu \ea computed volume-based scores and used a simple threshold for different classifications. Bron \ea used an SVM based on tissues density maps for many classification tasks.
A few methods based on deep learning have been also proposed. Ma \ea used a generative adversarial network (GAN) to classify structure volumes and Cth for 3-class differential diagnosis \cite{ma_differential_2020}. In \cite{hu_deep_2021}, Hu \ea used a Convolutional Neural Network (CNN) on intensities for multi-class classification.

Tables~\ref{table:sota_2_classes} and~\ref{table:sota_3_classes} respectively summarize the current performance of state-of-the-art methods proposed for binary classifications and 3-class classification. It should be noted that comparison has to be considered with caution. First, the data used for evaluation are not the same between methods. For instance, we use bvFTD patients (like \cite{moller_alzheimer_2016}) while others mixed several variants \cite{ma_differential_2020, hu_deep_2021, bron_multiparametric_2017, kim_machine_2019}. The metrics can also be different, some studies (\ie \cite{hu_deep_2021}) used ACC which is known to be sub-optimal when classes are unbalanced while other used BACC.

\begin{table}[ht]
    \centering
    \caption{Comparison of our method with current state-of-the-art methods for binary classification tasks. Our reported performances are the average of 10 repetitions and presented in \%. {\color{red} Red}: best result. The balanced accuracy (BACC) is used to assess the model performance except for \cite{hu_deep_2021} that used accuracy (ACC).}
    \label{table:sota_2_classes}
    \begin{tabular}{l@{\hskip 0.6em}c@{\hskip 0.6em}c@{\hskip 0.6em}c@{\hskip 0.6em}c}
        \hline
        Method  & \makecell{Dementia \\ diagnosis\\Dem. \vs CN} & \makecell{AD\\ diagnosis\\AD \vs CN}         & \makecell{FTD\\ diagnosis\\FTD \vs CN} & \makecell{Differential \\ diagnosis\\AD \vs FTD}\\
        \hline
        CNN on intensities \cite{hu_deep_2021} & - & 77.2 & 68.0 & 81.3\\
        LDA on Cth \cite{kim_machine_2019} & 86.2 & - & - & {\color{red} \textbf{89.8}}\\
        SVM on tissue map \cite{moller_alzheimer_2016} & - & 85.0 & 72.5 & 78.5\\
        Threshold on volumes \cite{Yu_mri-based_2021} & 85.7 & - & - & 83.0\\
        \hline
        Our method & {\color{red} \textbf{90.5}} & {\color{red} \textbf{90.3}}       & {\color{red} \textbf{93.3}} & 89.7\\
        
        \hline
    \end{tabular}
\end{table}

\begin{table}[ht]
    \centering
    \caption{Comparison of our method with current state-of-the-art methods for 3-class differential diagnosis AD \vs FTD \vs CN. {\color{red} Red}: best result. We denote ACC for accuracy, BACC for balanced accuracy and AUC for area under curve.}
    \label{table:sota_3_classes}
    \begin{tabular}{l@{\hskip 3em}c@{\hskip 2em}c@{\hskip 2em}c}
        \hline
        Method    & ACC         & BACC & AUC \\
        \hline
        SVM on tissue maps \cite{bron_multiparametric_2017}      & - & 75.0 & 90.0 \\
        GAN on Cth and volume \cite{ma_differential_2020}      & 88.3 & 85.3 & - \\
        CNN on intensities \cite{hu_deep_2021} & 66.8 & - & - \\
        \hline
        Our method & {\color{red} \textbf{90.4}} & {\color{red} \textbf{85.4}} & {\color{red} \textbf{95.3}}\\
        \hline
    \end{tabular}
\end{table}

For binary classification problems (see Table~\ref{table:sota_2_classes}), our method shows higher performance than other methods in specific disease diagnosis. For the 2-class differential diagnosis (AD \vs FTD) our result was comparable with the best one. For the 3-class differential diagnosis (see Table~\ref{table:sota_3_classes}), our method presented better performance in all metrics compared to other studies. Furthermore, contrary to other deep learning based methods, we provide an interpretable grading map allowing to localize abnormal regions in brain (see Section~\ref{section:interpretation}). These results highlight the potential value of our method in clinical practice. 

\subsection{Interpretation of deep grading map}
\label{section:interpretation}

\begin{figure}[ht]
\centering
\includegraphics[width=0.75\textwidth]{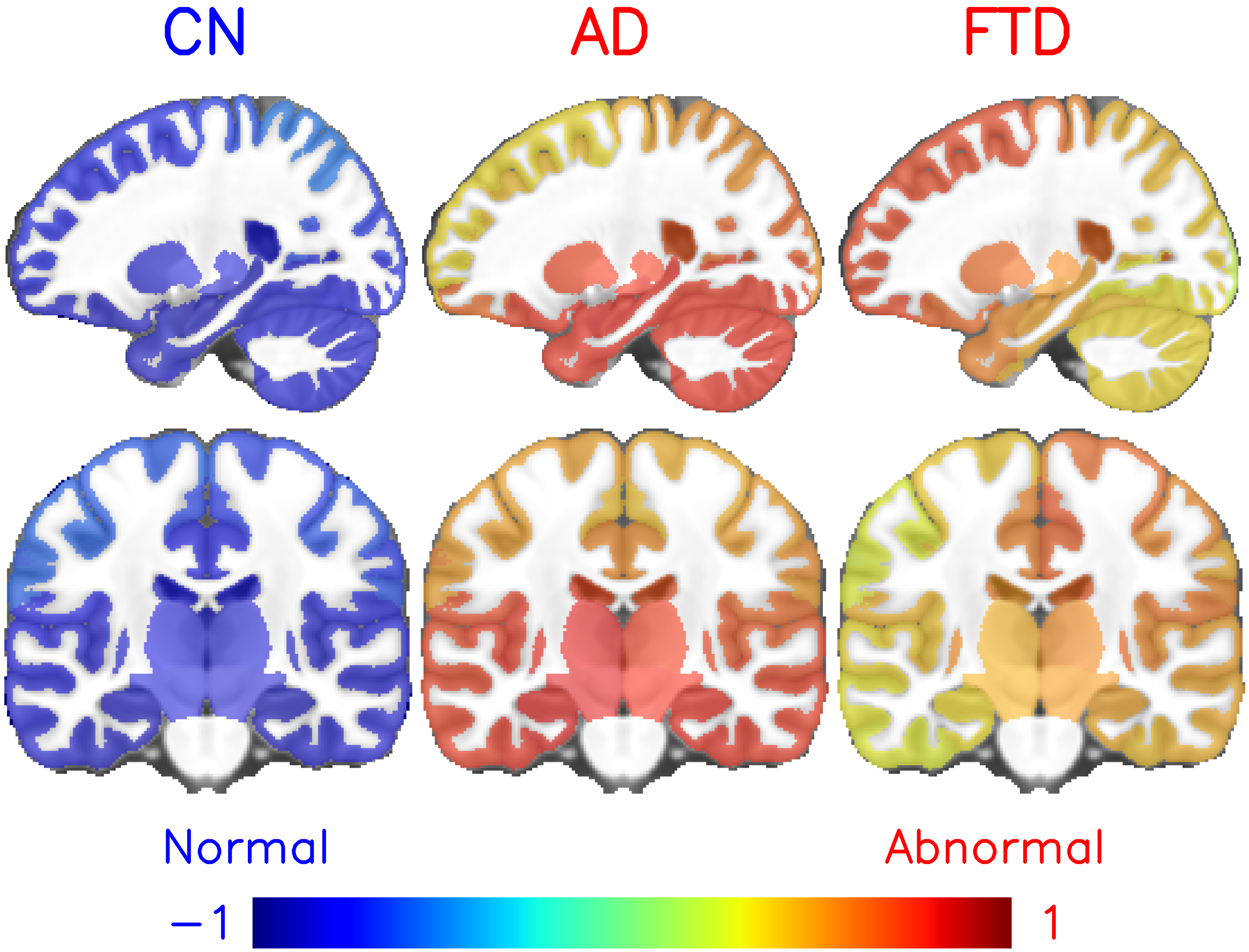}
\caption{Average grading map per group of subjects in the MNI space with neurological orientation (with the right of the patient at the right).}
\label{fig:two_views}
\end{figure}

To assess the interpretability of the deep grading map, we computed the average grading map for each group (\ie CN, AD and FTD). Figure~\ref{fig:two_views} shows sagittal and coronal views of these average grading maps. For the group of heathy people (\ie CN), all brain regions were detected as normal as expected. For the AD group, the average grading map highlighted regions around hippocampus. This area is well-known to be affected by AD \cite{schuff_mri_2009}. For FTD group, the abnormal regions were localized around the ventromedial frontal cortex. This region presents significant atrophy in FTD patients \cite{rosen_patterns_2002}.

We further investigated the top 10 structures with highest grading score in AD and FTD group. For AD group, eight over ten structures highlighted by the average grading map were known to be related to AD: left ventral diencephalon \cite{lebedeva_mri-based_2017}, left hippocampus \cite{frisoni_clinical_2010}, left amygdala and left inferior lateral ventricle \cite{coupe_lifespan_2019}, left parahippocampal gyrus and left entorhinal area \cite{kesslak_quantification_1991}, left thalamus \cite{de_jong_strongly_2008}, left basal forebrain \cite{whitehouse_alzheimers_1982}. For FTD group, we also found several structures related to the disease: bilateral anterior cingulate gyrus \cite{brambati_tensor_2007}, left medial frontal cortex and right middle frontal gyrus \cite{rosen_neuroanatomical_2005}, bilateral frontal pole \cite{wong_contrasting_2014}.

Moreover, the abnormality map of FTD appeared asymmetric between left and right hemisphere. This result is in line with our current knowledge on this disease \cite{harper_algorithmic_2014} since bvFTD patients usually exhibit higher atrophy in the right hemisphere.

\section{Conclusion}
In this paper, we proposed a new method for specific disease diagnosis (\ie AD \vs CN, FTD \vs CN) and differential diagnosis (\ie AD \vs FTD, AD \vs FTD \vs CN). Our purpose was to expand the knowledge about different dementia sub-types and offer an accurate diagnosis tool in a real clinical scenario. To this end, we extended a recent deep grading framework for multiple classification problems. First, we showed that our training strategy using multiple diseases (\ie AD and FTD) can improve the performance of specific disease diagnosis. Second, we proposed to combine the high capacity in disease detection offered by structure grading with the high accuracy on differential diagnosis provided by structure atrophy. As a result, our ensembling models achieved state-of-the-art in performance for various classification tasks.  Finally, the grading maps offer an easily interpretable tool to investigate dementia signatures over the entire brain. The structures highlighted by our grading map highly correlate with current physiopathological knowledge on AD and FTD. This presents a potential value of our method in clinical practice.

{\small
\subsubsection*{Acknowledgments}
This work benefited from the support of the project DeepvolBrain of the French National Research Agency (ANR-18-CE45-0013). This study was achieved within the context of the Laboratory of Excellence TRAIL ANR-10-LABX-57 for the BigDataBrain project. Moreover, we thank the Investments for the future Program IdEx Bordeaux (ANR-10-IDEX-03-02 and RRI "IMPACT"), the French Ministry of Education and Research, and the CNRS for DeepMultiBrain project.
}

\appendix

\bibliographystyle{splncs04}
\bibliography{paper2064.bib}

\begin{thebibliography}{10}
\providecommand{\url}[1]{\texttt{#1}}
\providecommand{\urlprefix}{URL }
\providecommand{\doi}[1]{https://doi.org/#1}

\bibitem{alladi_focal_2007}
Alladi, S., et~al.: Focal cortical presentations of {Alzheimer}'s disease.
  Brain  \textbf{130},  2636--2645 (2007)

\bibitem{avants_reproducible_2011}
Avants, B.B., et~al.: A reproducible evaluation of {ANTs} similarity metric
  performance in brain image registration. Neuroimage  \textbf{54},  2033--2044
  (2011)

\bibitem{bang_frontotemporal_2015}
Bang, J., other: Frontotemporal dementia. The Lancet  \textbf{386},  1672--1682
  (2015)

\bibitem{brambati_tensor_2007}
Brambati, S.M., et~al.: A tensor based morphometry study of longitudinal gray
  matter contraction in {FTD}. NeuroImage  \textbf{35},  998--1003 (2007)

\bibitem{bron_multiparametric_2017}
Bron, E.E., et~al.: Multiparametric computer-aided differential diagnosis of
  {Alzheimer}’s disease and frontotemporal dementia using structural and
  advanced {MRI}. European Radiology  \textbf{27},  3372--3382 (2017)

\bibitem{coupe_lifespan_2019}
Coupé, P., et~al.: Lifespan {Changes} of the {Hum} {Brain} {In}
  {Alzheimer}’s {Disease}. Sci Rep  \textbf{9}, ~3998 (2019)

\bibitem{coupe_assemblynet_2020}
Coupé, P., et~al.: {AssemblyNet}: {A} large ensemble of {CNNs} for {3D} whole
  brain {MRI} segmentation. NeuroImage  \textbf{219},  117026 (2020)

\bibitem{davatzikos_individual_2008}
Davatzikos, C., et~al.: Individual patient diagnosis of {AD} and {FTD} via
  high-dimensional pattern classification of {MRI}. NeuroImage  \textbf{41},
  1220--1227 (2008)

\bibitem{du_different_2006}
Du, A.T., et~al.: Different regional patterns of cortical thinning in
  {Alzheimer}'s disease and frontotemporal dementia. Brain  \textbf{130},
  1159--1166 (2006)

\bibitem{ellis_australian_2009}
Ellis, K.A., et~al.: The {Australi} {Imaging}, {Biomarkers} and {Lifestyle}
  ({AIBL}) study of aging: methodology and baseline characteristics of 1112
  individuals recruited for a longitudinal study of {Alzheimer}'s disease. Int
  Psychogeriatr  \textbf{21},  672--687 (2009)

\bibitem{frisoni_clinical_2010}
Frisoni, G.B., et~al.: The clinical use of structural {MRI} in {Alzheimer}
  disease. Nat Rev Neurol  \textbf{6},  67--77 (2010)

\bibitem{harper_algorithmic_2014}
Harper, L., et~al.: An algorithmic approach to structural imaging in dementia.
  Journal of Neurology, Neurosurgery \& Psychiatry  \textbf{85},  692--698
  (2014)

\bibitem{hu_deep_2021}
Hu, J., et~al.: Deep {Learning}-{Based} {Classification} and {Voxel}-{Based}
  {Visualization} of {Frontotemporal} {Dementia} and {Alzheimer}’s {Disease}.
  Frontiers in Neuroscience  \textbf{14},  626154 (2021)

\bibitem{hutchinson_neuropsychological_2007}
Hutchinson, A.D., et~al.: Neuropsychological deficits in frontotemporal
  dementia and {Alzheimer}'s disease: a meta-analytic review. Journal of
  Neurology, Neurosurgery, and Psychiatry  \textbf{78},  917--928 (2007)

\bibitem{jack_alzheimers_2008}
Jack, C.R., et~al.: The {Alzheimer}'s {Disease} {Neuroimaging} {Initiative}
  ({ADNI}): {MRI} methods. J Magn Reson Imaging  \textbf{27},  685--691 (2008)

\bibitem{de_jong_strongly_2008}
de~Jong, L.W., et~al.: Strongly reduced volumes of putamen and thalamus in
  {Alzheimer}'s disease: an {MRI} study. Brain  \textbf{131},  3277--3285
  (2008)

\bibitem{kesslak_quantification_1991}
Kesslak, J.P., et~al.: Quantification of magnetic resonance scans for
  hippocampal and parahippocampal atrophy in {Alzheimer}'s disease. Neurology
  \textbf{41},  51--54 (1991)

\bibitem{kim_machine_2019}
Kim, J.P., et~al.: Machine learning based hierarchical classification of
  frontotemporal dementia and {Alzheimer}'s disease. NeuroImage: Clinical
  \textbf{23},  101811 (2019)

\bibitem{kipf_semi-supervised_2017}
Kipf, T.N., et~al.: Semi-{Supervised} {Classification} with {Graph}
  {Convolutional} {Networks}. In: 5th International Conference on Learning
  Representations, {ICLR} 2017 (2017)

\bibitem{lamontagne_oasis_3_2019}
LaMontagne, P.J., et~al.: {OASIS}-3: {Longitudinal} {Neuroimaging}, {Clinical},
  and {Cognitive} {Dataset} for {Normal} {Aging} and {Alzheimer} {Disease}.
  preprint, Radiology and Imaging (2019)

\bibitem{lebedeva_mri-based_2017}
Lebedeva, A.K., et~al.: {MRI}-{Based} {Classification} {Models} in {Prediction}
  of {Mild} {Cognitive} {Impairment} and {Dementia} in {Late}-{Life}
  {Depression}. Front Aging Neurosci  \textbf{9}, ~13 (2017)

\bibitem{ma_differential_2020}
Ma, D., et~al.: Differential {Diagnosis} of {Frontotemporal} {Dementia},
  {Alzheimer}'s {Disease}, and {Normal} {Aging} {Using} a {Multi}-{Scale}
  {Multi}-{Type} {Feature} {Generative} {Adversarial} {Deep} {Neural} {Network}
  on {Structural} {Magnetic} {Resonance} {Images}. Frontiers in Neuroscience
  \textbf{14}, ~853 (2020)

\bibitem{malone_miriadpublic_2013}
Malone, I.B., et~al.: {MIRIAD}—{Public} release of a multiple time point
  {Alzheimer}'s {MR} imaging dataset. NeuroImage  \textbf{70},  33--36 (2013)

\bibitem{manjon_robust_2008}
Manjón, J.V., et~al.: Robust {MRI} brain tissue parameter estimation by
  multistage outlier rejection. Magn Reson Med  \textbf{59},  866--873 (2008)

\bibitem{manjon_adaptive_2010}
Manjón, J.V., et~al.: Adaptive non-local means denoising of {MR} images with
  spatially varying noise levels. J Magn Reson Imaging  \textbf{31},  192--203
  (2010)

\bibitem{manjon_nonlocal_2014}
Manjón, J.V., et~al.: Nonlocal intracranial cavity extraction. Int J Biomed
  Imaging  \textbf{2014},  820205 (2014)

\bibitem{mckhann_diagnosis_2011}
McKhann, G.M., et~al.: The diagnosis of dementia due to {Alzheimer}'s disease:
  recommendations from the {National} {Institute} on {Aging}-{Alzheimer}'s
  {Association} workgroups on diagnostic guidelines for {Alzheimer}'s disease.
  Alzheimer's \& Dementia: The Journal of the Alzheimer's Association
  \textbf{7},  263--269 (2011)

\bibitem{moller_alzheimer_2016}
Möller, C., et~al.: Alzheimer {Disease} and {Behavioral} {Variant}
  {Frontotemporal} {Dementia}: {Automatic} {Classification} {Based} on
  {Cortical} {Atrophy} for {Single}-{Subject} {Diagnosis}. Radiology
  \textbf{279},  838--848 (2016)

\bibitem{neary_frontotemporal_2005}
Neary, D., et~al.: Frontotemporal dementia. The Lancet Neurology  \textbf{4},
  771--780 (2005)

\bibitem{nguyen_deep_2021}
Nguyen, H.D., et~al.: Deep {Grading} {Based} on {Collective} {Artificial}
  {Intelligence} for {AD} {Diagnosis} and {Prognosis}. In: Interpretability of
  {Machine} {Intelligence} in {Medical} {Image} {Computing}, and {Topological}
  {Data} {Analysis} and {Its} {Applications} for {Medical} {Data}. vol. 12929,
  pp. 24--33 (2021)

\bibitem{pedregosa11a}
Pedregosa, F., other: Scikit-learn: Machine learning in python. Journal of
  Machine Learning Research  \textbf{12},  2825--2830 (2011)

\bibitem{rabinovici_distinct_2008}
Rabinovici, G., et~al.: Distinct {MRI} {Atrophy} {Patterns} in
  {Autopsy}-{Proven} {Alzheimer}'s {Disease} and {Frontotemporal} {Lobar}
  {Degeneration}. American Journal of Alzheimer's Disease \& Other Dementiasr
  \textbf{22},  474--488 (2008)

\bibitem{rascovsky_sensitivity_2011}
Rascovsky, K., et~al.: Sensitivity of revised diagnostic criteria for the
  behavioural variant of frontotemporal dementia. Brain  \textbf{134},
  2456--2477 (2011)

\bibitem{rosen_patterns_2002}
Rosen, H.J., et~al.: Patterns of brain atrophy in frontotemporal dementia and
  semantic dementia. Neurology  \textbf{58},  198--208 (2002)

\bibitem{rosen_neuroanatomical_2005}
Rosen, H.J., et~al.: Neuroanatomical correlates of behavioural disorders in
  dementia. Brain  \textbf{128},  2612--2625 (2005)

\bibitem{schuff_mri_2009}
Schuff, N., et~al.: {MRI} of hippocampal volume loss in early {Alzheimer}'s
  disease in relation to {ApoE} genotype and biomarkers. Brain  \textbf{132},
  1067--1077 (2009)

\bibitem{tustison_n4itk_2010}
Tustison, N.J., et~al.: {N4ITK}: improved {N3} bias correction. IEEE Trans Med
  Imaging  \textbf{29},  1310--1320 (2010)

\bibitem{whitehouse_alzheimers_1982}
Whitehouse, P., et~al.: Alzheimer's {Disease} and {Senile} {Dementia}: {Loss}
  of {Neurons} in the {Basal} {Forebrain}. Science  \textbf{215},  1237--1239
  (1982)

\bibitem{wong_contrasting_2014}
Wong, S., et~al.: Contrasting {Prefrontal} {Cortex} {Contributions} to
  {Episodic} {Memory} {Dysfunction} in {Behavioural} {Variant} {Frontotemporal}
  {Dementia} and {Alzheimer}’s {Disease}. PLoS ONE  \textbf{9},  e87778
  (2014)

\bibitem{yew_lost_2013}
Yew, B., et~al.: Lost and forgotten? {Orientation} versus memory in
  {Alzheimer}'s disease and frontotemporal dementia. Journal of Alzheimer's
  disease: JAD  \textbf{33},  473--481 (2013)

\bibitem{Yu_mri-based_2021}
Yu, Q., et~al.: An {MRI}-based strategy for differentiation of frontotemporal
  dementia and {Alzheimer}’s disease. Alzheimer's Research \& Therapy
  \textbf{13}, ~23 (2021)

\end{thebibliography}

\end{document}